\documentclass[aps,pra,reprint,showpacs,floatfix]{revtex4-1}

\usepackage{amsmath}
\usepackage{slashed}
\usepackage{txfonts}
\usepackage{microtype}
\usepackage{graphicx}
\usepackage[
breaklinks=true
]{hyperref}
\usepackage{color}
\usepackage{isomath}
\def\sout{\bgroup\markoverwith
{\textcolor{red}{\rule[0.5ex]{2pt}{0.5pt}}}\ULon}
\def\be{\begin{equation}}
\def\ee{\end{equation}}
\def\bes{\begin{equation*}}
\def\ees{\end{equation*}}
\def\bea{\begin{eqnarray}}
\def\eea{\end{eqnarray}}
\def\beas{\begin{eqnarray*}}
\def\eeas{\end{eqnarray*}}
\def\bal#1\eal{\begin{align}#1\end{align}}
\def\bals#1\eals{\begin{align*}#1\end{align*}}
\newcommand{\bra}[1]{\langle #1|}
\newcommand{\ket}[1]{|#1\rangle}

\renewcommand{\vec}[1]{\mathbf{#1}} 
\newcommand{\del}{\partial}

\graphicspath{{figures/}}

\renewcommand*{\vec}[1]{\boldsymbol{#1}}

\begin{document}
\title{The Wigner time delay for laser induced tunnel-ionization via the electron propagator}

\author{Enderalp Yakaboylu}
\author{Michael Klaiber}
\author{Karen Z. Hatsagortsyan}
\affiliation{Max-Planck-Institut f{\"u}r Kernphysik, Saupfercheckweg~1,
  69117~Heidelberg, Germany}
\date{\today}

\begin{abstract}

Recent attoclock experiments using the attsecond angular streaking technique enabled the
measurement of the tunneling time delay during laser induced strong field ionization. Theoretically
the tunneling time delay is commonly modelled by the Wigner time delay concept which is derived from
the derivative of the electron wave function phase with respect to energy. Here, we present an
alternative method for the calculation of the Wigner time delay by using the fixed energy
propagator. The developed formalism is applied to the nonrelativistic as well as to the relativistic
regime of the tunnel-ionization process from a zero-range potential, where in the latter regime the
propagator can be given by means of the proper-time method.

\end{abstract}

\maketitle

\section{Introduction}

The tunneling of an electron through a potential barrier is a typical quantum mechanical phenomenon.
It has been in the focus of both theoretical and experimental attention since the formulation of
quantum mechanics (see, e.g., \cite{Razavy_2003} for a comprehensive review). In particular the
issue of whether the motion of a particle under a barrier is instantaneous or not is a long standing
and controversial problem in physics since MacColl's first attempt to consider it in 1932
\cite{MacColl_1932}. Recent interest to this problem has been renewed by a unique opportunity
offered by attosecond angular streaking techniques for measuring the tunneling time delay during
laser-induced tunnel-ionization \cite{Eckle_2008a,Eckle_2008b,Pfeiffer_2012a,Landsman_2013}. This
novel experimental technique offers the measurement of tunneling time with unprecedented resolution of
tens of attoseconds \cite{atto_2013}. 

For the generic problem of the tunneling time delay, different definitions have been proposed and
the discussion of their relevance still continues
\cite{Eisenbud_1948,Wigner_1955,Smith_1960,Landauer_1994,Sokolovski_2008,Steiberg_2008,Ban_2010,
Galapon_2012,Vona_2013}. 
The Keldysh time introduced in the strong field ionization theory \cite{Keldysh_1965}
describes the
formation time of the ionization process \cite{Brabec_2013} (see also Fig.~11 in
\cite{Yakaboylu_2013_rt}) and is not relevant to the attoclock time delay measurement
\cite{Eckle_2008a,Eckle_2008b,Pfeiffer_2012a,Landsman_2013}. One of the most accepted definitions
for the tunneling time delay is the
Wigner time delay concept \cite{Eisenbud_1948,Wigner_1955,Smith_1960}. This concept is based on the time difference 
between the quasiclassical and the Wigner trajectory, which is the trajectory of the peak of the wave packet, at a 
remote distance. A mathematical definition of the Wigner trajectory can be given via the derivative of the
phase of the steady-state wave function of the tunneling particle. Namely, the time is expressed via
the derivative with respect to energy and the coordinate via the derivative with respect to the
corresponding momentum.

Recently, an intuitive picture of the relativistic regime of the laser-induced strong field
ionization was developed in \cite{Klaiber_2013c,Yakaboylu_2013_rt}. It was shown that the tunneling
picture applies also in the relativistic regime with a modification that the energy levels becomes 
position dependent. This modification accounts for the electron kinetic energy change during the
tunneling connected with the electron motion along the laser field due to the effect of the magnetic
field induced Lorentz force. The tunneling time delay was investigated in \cite{Yakaboylu_2013_rt}
for the tunnel-ionization process by adapting Wigner's approach for tunnel-ionization. In
the relativistic regime of tunneling the Wigner trajectory under-the-barrier is shifted along the
laser propagation direction due to the magnetically induced Lorentz force, which can be observed
physically as a shift of the peak of the tunneled out electron wave packet at the exit of the
barrier. In \cite{Yakaboylu_2013_rt} the Wigner trajectory is calculated explicitly extracting the
phase of the wave function which is a rather cumbersome procedure especially in the relativistic
regime.

In this brief report, we present an alternative method for the calculation of the Wigner trajectory
in terms of the phase of the fixed energy propagator (the Green function of the time-independent
Schr\"{o}dinger equation). This method provides an easier way to calculate the tunneling time delay
and the Wigner trajectory in particular cases. We apply the developed formalism to the
nonrelativistic regime of tunnel-ionization from a zero-range atomic potential under the effect of a
constant and uniform electric field, and the relativistic regime under the effect of a constant and
uniform crossed field. The latter field configuration corresponds to the relativistic quasistatic
regime of strong field ionization when the Keldysh parameter $\gamma=\omega \sqrt{2I_p}/E_0$ is
small ($\gamma \ll 1$) \cite{Keldysh_1965}, where $I_p$ is the ionization potential, $\omega$ and
$E_0$ are the laser frequency and the field amplitude, respectively.

Atomic units (a.~u.) and the metric convention $g = (+,-,-,-)$ are used throughout the paper.

\section{The phase of the fixed energy propagator}

Our fundamental definition of the Wigner trajectory follows from the relation between the spacetime
propagator
\be
K(x,x';t.t') = \bra{x} U(t,t') \ket{x'} 
\ee
and the fixed energy propagator 
\be
G(x,x';\varepsilon) = \bra{x} \frac{1}{\varepsilon - H + i \epsilon_F} \ket{x'} \, ,
\ee
with energy $\varepsilon$, the Hamiltonian $H$, the time evolution operator $U(t,t')$ and the
standard Feynman $i \epsilon_F$ prescription \cite{Weinberg_1995}. First, the spacetime propagator
can be written as
\be
K(x,x';t) = \frac{1}{2 \pi} \int_{-\infty}^\infty d \varepsilon \, \exp\left( - i \varepsilon t
\right) G(x,x';\varepsilon) \, ,
\ee
where we set the initial time zero, $t' = 0$. Then, the fixed energy propagator is split into its
phase $\phi$ and its amplitude $A$ such that the spacetime propagator reads
\be
\label{propagator}
K(x,x';t) = \frac{1}{2 \pi} \int_{-\infty}^\infty d \varepsilon \, A(x,x', \varepsilon)
\exp\left( - i \varepsilon t  + i \phi(x,x', \varepsilon) \right)  \, .
\ee

The propagator~(\ref{propagator}) connects the space points $x$ and $x'$ in a time interval
$t$ and
its value will be maximal when the phase of the fixed energy propagator $\phi(x,x', \varepsilon)$
fulfills the stationary phase condition
\be
\label{wigner_traject_via_fix_energy_prop_fundamental_0}
t - \frac{\del \phi(x,x', \varepsilon)}{\del \varepsilon} = 0 \, .
\ee
Then a trajectory which fulfills condition~(\ref{wigner_traject_via_fix_energy_prop_fundamental_0})
with a certain energy of the incoming wave packet $\varepsilon_0$ maximizes the propagator. In
other words, the trajectory given by the relation
\be
\label{wigner_traject_via_fix_energy_prop_fundamental}
t^w (x,x') = \left. \frac{\del \phi(x,x', \varepsilon)}{\del
\varepsilon} \right|_{\varepsilon = \varepsilon_0} \,
\ee
traces the maximal of the propagator. In fact, the
trajectory~(\ref{wigner_traject_via_fix_energy_prop_fundamental}) is nothing but the 
definition of the so-called Wigner trajectory \cite{Eisenbud_1948,Wigner_1955,Smith_1960}.
Here we should stress that the
identification of the Wigner trajectory by means of the phase of the electron's propagator is
similar to the procedure given in \cite{Yakaboylu_2013_rt}. Nevertheless,
Eq.~(\ref{wigner_traject_via_fix_energy_prop_fundamental}) is based on a more fundamental approach
that prevents us to discuss the appropriate conditions in order to identify a well-defined Wigner
trajectory by means of wave packets elaborated in \cite{Yakaboylu_2013_rt}. 

In the tunnel-ionization case one can argue that the quasiclassical trajectory and the Wigner
trajectory have to coincide at the entry point of the tunneling barrier $x_i$
\cite{Yakaboylu_2013_rt}. Therefore, imposing this condition, we obtain the Wigner
trajectory for the tunnel-ionization in terms of the phase of the fixed energy propagator as
\be
\label{wigner_traject_via_fix_energy_prop}
t^w_\mathrm{TI} (x) \equiv \left. \frac{\del \phi(x,x_i, \varepsilon)}{\del
\varepsilon} \right|_{\varepsilon = \varepsilon_0} - \left. \frac{\del \phi(x_i,x_i,
\varepsilon)}{\del \varepsilon} \right|_{\varepsilon = \varepsilon_0}\, .
\ee
where $\varepsilon_0 = - I_p$ for the nonrelativistic regime of tunnel-ionization whereas $\varepsilon_0 = c^2 - I_p$ 
for the relativistic regime with the ionization energy of the ground state of the H-like ion $I_p = c^2 
- \sqrt{c^4 - c^2 \kappa^2}$, atomic momentum $\kappa$ and speed of light $c$.

The phase of the fixed energy propagator can be inferred by the inverse Fourier transform of the
spacetime propagator as
\be
\label{inverse_FT}
G(x,x';\varepsilon) = \int_0^\infty dt \, \exp\left(i \varepsilon t \right) 
K(x,x';t) \, .
\ee
Moreover, if the fixed energy propagator~(\ref{inverse_FT}) is calculated by the saddle
point
approximation, then the Wigner trajectory~(\ref{wigner_traject_via_fix_energy_prop}) coincides with
the entire quasiclassical trajectory $t^c (x)$, which is instantaneous under-the-barrier and obeys
Newton's law outside the tunneling barrier.

Finally, we can define the Wigner time delay, which measures the time difference between
quasiclassical trajectory and the Wigner trajectory at a fixed space point far away from the barrier, as
\be
\tau^w = t^w_\mathrm{TI}(\infty) - t^c (\infty) \, .
\ee

\subsection{Nonrelativistic case: in a constant and uniform electric field}

\begin{figure}
  \centering
  \includegraphics[width=0.8\linewidth]{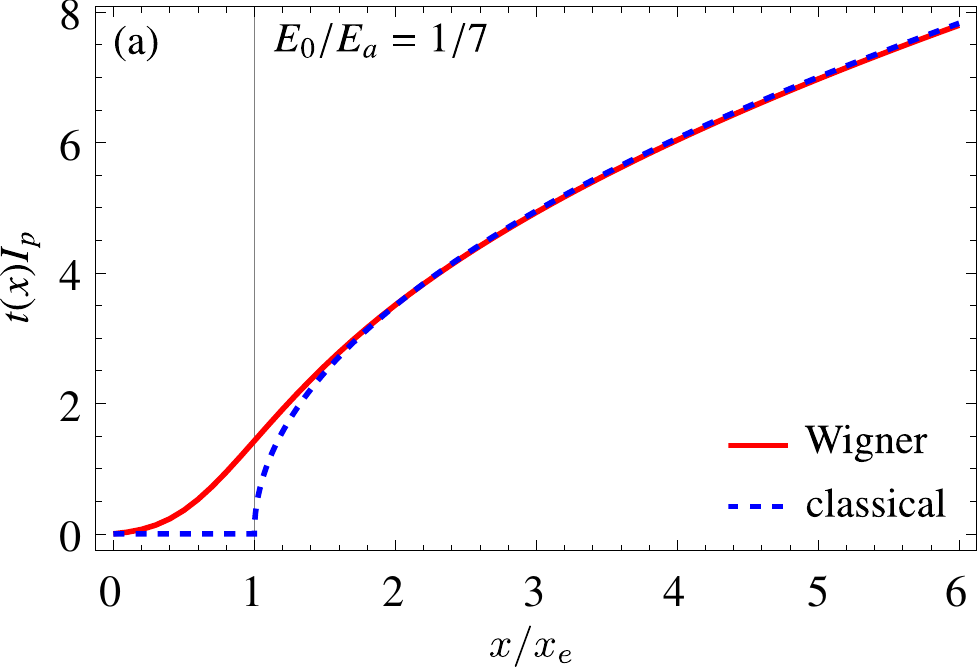}\\[3ex]
  \includegraphics[width=0.8\linewidth]{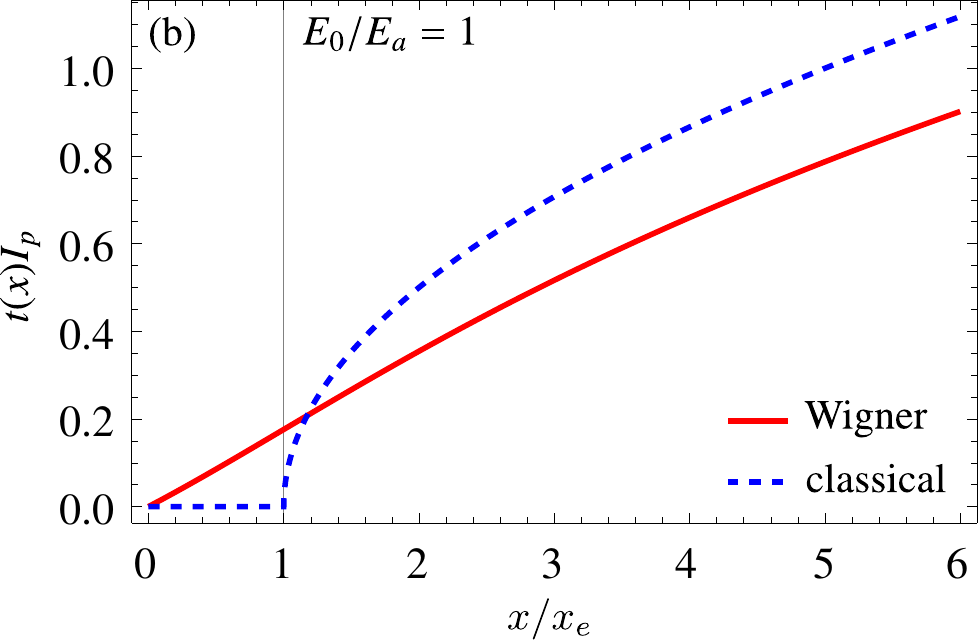}
  \caption{(Color online) Comparison of the Wigner trajectory (red solid
    line) and the classical trajectory (blue dashed line) for nonrelativistic tunnel
    ionization from a zero-range potential under the effect of a constant and uniform electric
    field; the deep-tunneling regime (a) with $E_0/E_a=1/7$ and the
    near-threshold-tunneling regime (b) with $E_0/E_a=1$. The vertical
    black line indicates the exit coordinate and the applied parameter is $\kappa =1$.}
  \label{time_delay_e_field_fixpro}
\end{figure}

Let us apply the developed formalism first in the nonrelativistic regime of tunnel-ionization from a
zero-range potential by neglecting the magnetic field of the laser and considering solely the
electric field component
\be
\label{cons_uni_e_field}
\vec{E} = E_0 \,  \hat{\vec{x}} \, .
\ee


The exact spacetime propagator for those particles whose Lagrangian is a quadratic function
of the
coordinate and the velocity coincides with the quasiclassical propagator, i.e., the
classical path dominates the Feynman path integral \cite{Feynman_Hibbs,Schulman,Kleinert}. In this
case, the propagator can be given by
\be
K (x,x';t) = \sqrt{\frac{-\del^2 S_c}{(2 \pi i)  \, \del x \del x'}} \exp\left( i \, S_c
(x,x',t)\right) 
\ee 
with the classical action $S_c$, which is the action evaluated along the classical trajectory.
Then, the spacetime propagator along the electric field direction-$x$ in the length gauge
$A^\mu = (- x E_0, 0)$ can be written as
\be
K(x,0;t) = \sqrt{\frac{1}{2 \pi i \, t}} \exp \left(\frac{i x^2}{2 t} + \frac{i E_0 t x}{2} -
\frac{i E_0^2 t^3 }{24}\right) \, .
\ee 
Here the tunneling entry point for the tunnel-ionization form a zero-range potential can be set as
$x_i = 0$. Then using the inverse Fourier transform~(\ref{inverse_FT}) and the
definition~(\ref{wigner_traject_via_fix_energy_prop}), we obtain the Wigner trajectory, see
Fig.~\ref{time_delay_e_field_fixpro}.

The classical trajectory, on the other hand, can be governed by
\be
\label{class_traject_e_field_t}
x(t) = x_e + \frac{1}{2} E_0 t^2
\ee
where the initial velocity at the tunnel exit can be set to zero and the tunnel exit point is given 
by $x_e = I_p /E_0$.
 
Now, we can compare the classical trajectory (\ref{class_traject_e_field_t}) with the Wigner
trajectory (\ref{wigner_traject_via_fix_energy_prop}) for nonrelativistic tunnel-ionization from
a zero-range potential. We did a comparison for two sets of parameters: First for $E_0 / E_a = 1/7$,
$\kappa = 1$ with the atomic field $E_a = \kappa^3$, corresponding to the deep-tunneling regime, and
for $E_0 / E_a = 1$, $\kappa = 1$, which represents the near-threshold-tunneling regime of the
tunnel-ionization \cite{Yakaboylu_2013_rt}. Fig.~\ref{time_delay_e_field_fixpro} demonstrates that
for the deep-tunneling regime the Wigner time delay vanishes, while for the near-threshold-tunneling
regime it persists and is detectable at remote distances. The results are consistent with
Ref.~\cite{Yakaboylu_2013_rt}.

\begin{figure}
  \centering
  \includegraphics[width=0.8\linewidth]{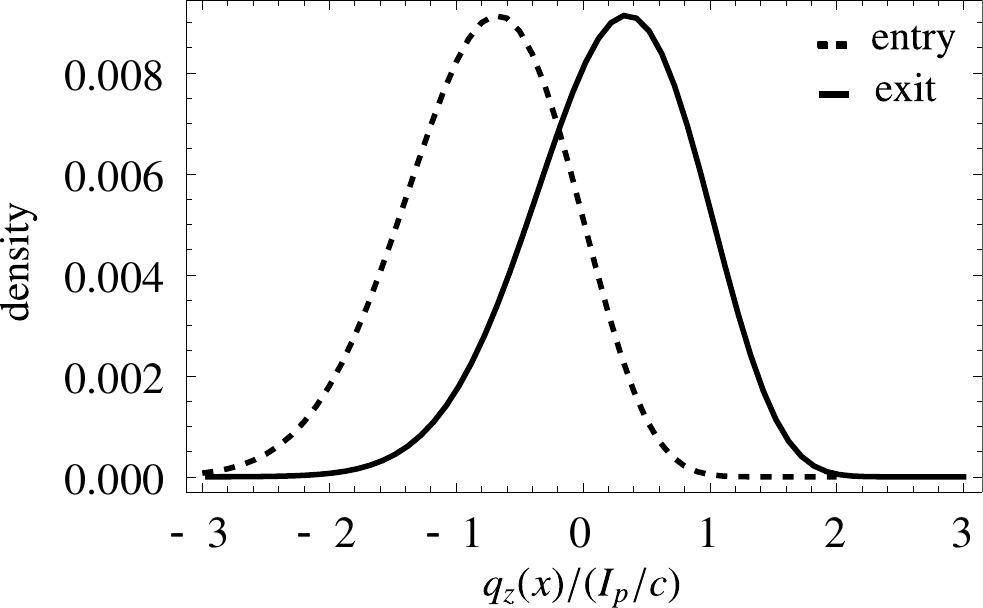}
  \caption{Tunneling probability vs. the kinetic momentum along
  the propagation direction of the crossed field at the tunnel entry (dashed line) and
  at the tunnel exit (solid line). In the case of tunnel-ionization from a zero-range potential, the
  values are independent from the barrier suppression parameter $E_0/E_a$ and the transversal
  momentum transfer is $I_p/c$. The applied parameters for the figure are $E_0/E_a=1/7$ and $\kappa
  =90$.}
  \label{mom_kick_cross_field}
\end{figure}

\subsection{Relativistic case: in a constant and uniform crossed field}



\begin{figure}
  \centering
  \includegraphics[width=0.8\linewidth]{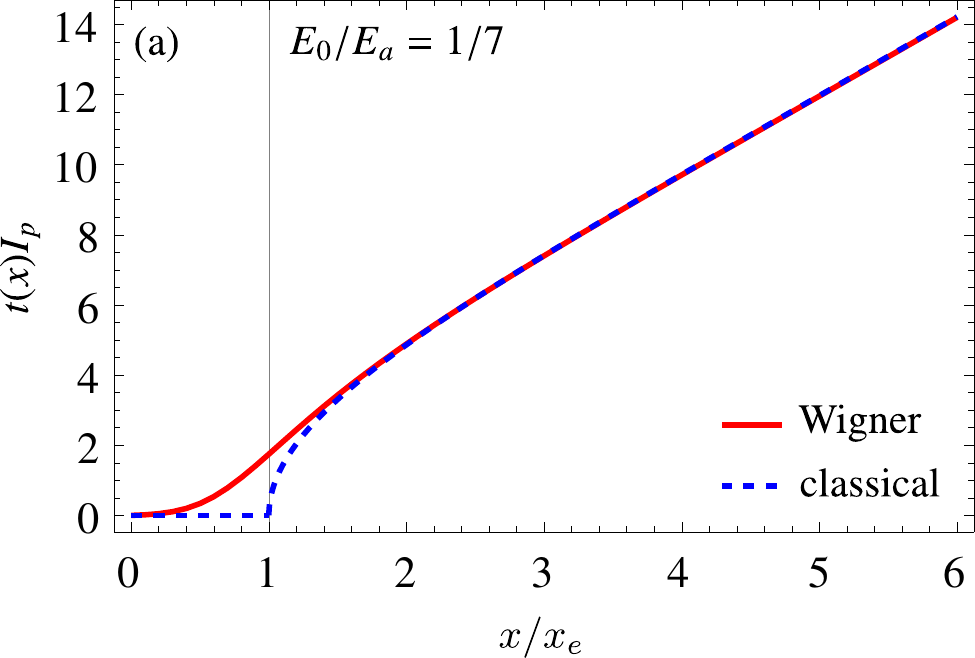}\\[3ex]
  \includegraphics[width=0.8\linewidth]{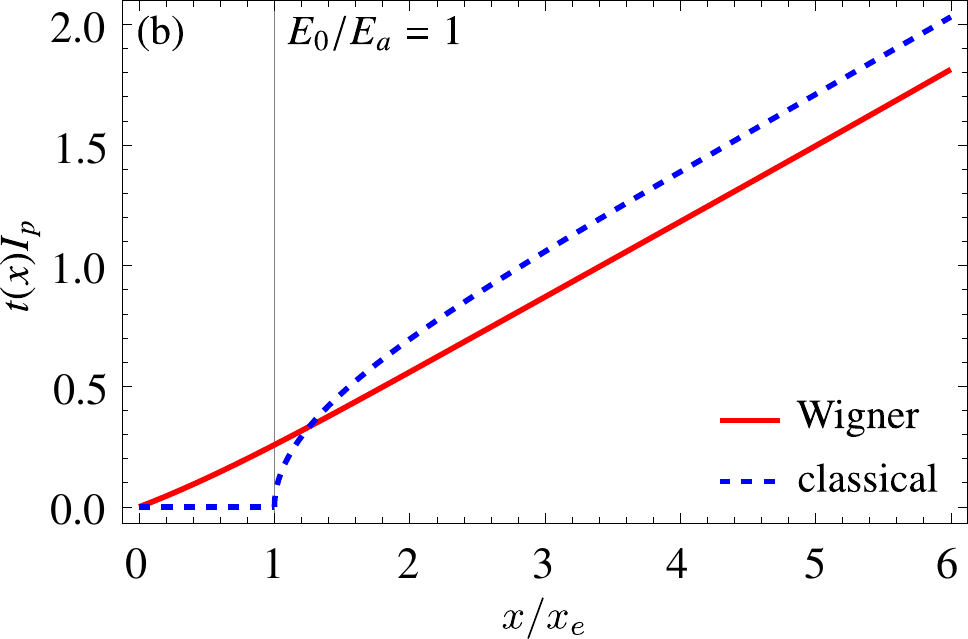}
  \caption{(Color online) Comparison of the Wigner trajectory (red solid line) and the classical
  trajectory (blue dashed line) for relativistic tunnel-ionization from a zero-range potential under
  the effect of a constant and uniform crossed field; the deep-tunneling regime (a) with
  $E_0/E_a=1/7$ and the near-threshold-tunneling regime (b) with $E_0/E_a=1$. The vertical black
  line indicates the exit coordinate and the applied parameter is $\kappa =90$.}
\label{time_delay_cross_fields_fixpro}
\end{figure}

Next, the relativistic tunnel-ionization from a zero-range potential is considered when the magnetic
field component of laser cannot be neglected and the laser field can be approximated by a constant
and uniform crossed field
\bal
\label{crossed_fields} 
\vec{E} & = E_0 \, \hat{\vec{x}} \, , \\ 
\nonumber \vec{B} & = E_0 \, \hat{\vec{y}} \, .
\eal

\pagebreak

The (3+1) dimensional relativistic spacetime propagator can be calculated within the proper time
method 
\cite{Fock_1937, 
Feynman_1950,Schwinger_1951,Dirac_qm,Fradkin_1991,Brink_1977,Polyakov_st}. Then, the corresponding
spacetime propagator for the crossed field~(\ref{crossed_fields}) in the G{\"o}ppert-Mayer gauge 
$A^\mu = - x E_0 \, k^\mu$ can be written as (see \cite{Yakaboylu_2013_pro})
\bal
\nonumber & K(x^\mu,0) = -\frac{i}{2} \exp\left(\frac{i E_0}{2 c} \epsilon \cdot x \, k \cdot x
\right) \int_0^\infty d \tau \, \frac{1}{(2 \pi \tau)^2}  \\
& \times \exp\left(-i \frac{x \cdot x}{2 \tau}-\frac{i \tau}{24 c^2}x \cdot F \cdot F \cdot x-\frac{i c^2
\tau}{2} \right) \,
\eal
with the wave vector $k^\mu = (1,0,0,1)$ and the polarization vector $\epsilon^\mu = (0,1,0,0)$, where we set the 
initial spacetime point $x'^\mu=0$ similar to the nonrelativistic case.

The fixed energy and the fixed transversal momenta propagator along the laser's polarization
direction can be calculated via the Fourier transform
\be
G(x,0;\varepsilon, p_z, p_y) = \int d t \, d z \, d y \, K(x^\mu,0) \exp\left( i \varepsilon t - i
p_z z - p_y y \right) \, ,
\ee  
which can be given by
\begin{widetext}
\be
\label{fixed_energy_prop_cross_fields} G(x;\varepsilon, p_z, p_y)= -\frac{(-1)^{3/4}}{2 c}
\int_0^\infty d \tau \, \frac{1}{\sqrt{2 \pi \, \tau}} \exp \left(\frac{i x^2 }{2 \tau} -\frac{i
\tau\left(c^2+p_y^2+p_z^2\right)}{2} + \frac{i
\tau (c E_0 p_z x+\varepsilon (\varepsilon -E_0 x))}{2 c^2} - \frac{i
 \tau ^3 E_0^2 (\varepsilon -c p_z)^2}{24 c^4} \right) \, .
\ee
\end{widetext}

In order to be able to plot the Wigner trajectory, we have to define $p_y$ as well as $p_z$
in
Eq.~(\ref{fixed_energy_prop_cross_fields}). Firstly, one can set $p_y = 0$ without loss of
generality for the maximal tunneling probability. However, in contrast to the nonrelativistic
tunnel-ionization process, the existence of the magnetic field leads to a non-zero value for $p_z$
for
the maximal tunneling probability, which is a consequence of the fact that there is momentum
transfer along the propagation direction of the laser in the relativistic regime
\cite{Klaiber_2013c,Yakaboylu_2013_rt}. This will be deduced in the following. The
tunneling probability for a given energy $\varepsilon_0 = c^2 - I_p$ can be given in terms of the
propagator as
\be
\label{trans_prob_cross_field}
\left| T \right|^2 = \frac{\left| G(x_e;\varepsilon_0, p_z) \right|^2}{\left|
G(0;\varepsilon_0, p_z) \right|^2} \, ,
\ee
where the quasiclassical tunnel exit point $x_e$ can be calculated via the condition
\be
\left(c^2 - I_p - x_e E_0 \right)^2 = c^2 \left(p_z - x_e \frac{E_0}{c} \right)^2 + c^4 \, ,
\ee
which yields
\be
\label{tunnel_exit_crossed_field}
x_e = \frac{I_p^2 - c^2 (2 I_p + p_z^2)}{2 E_0 (c^2 - I_p - c p_z)} \, .
\ee
The transition probability~(\ref{trans_prob_cross_field}) reveals the most probable
tunneling probability for a certain transversal momentum $p_z$, see Fig.~\ref{mom_kick_cross_field}.
This indicates that during the tunneling there is a momentum transfer along the propagation
direction of the crossed field. The kinetic momentum $q_z (x) = p_z - x E_0 /c$ with the maximal
tunneling probability at the tunneling entry is $q_z (0) = p_z \sim - 2 I_p /(3 c)$, whereas at the
exit it is $q_z (x_e) \sim I_p /(3 c) $. As a consequence, the momentum transfer along the laser's
propagation direction is $I_p/c$. Furthermore, in contrast to tunnel-ionization from a Coulomb
potential case, the momenta at the entry and the exit, and hence the momentum transfer are
independent from the barrier suppression parameter $E_0/E_a$ for a zero-range potential. (see 
\cite{Klaiber_2013c,Yakaboylu_2013_rt} for further details).

\begin{figure}[h!] 
  \centering
  \includegraphics[width=0.8\linewidth]{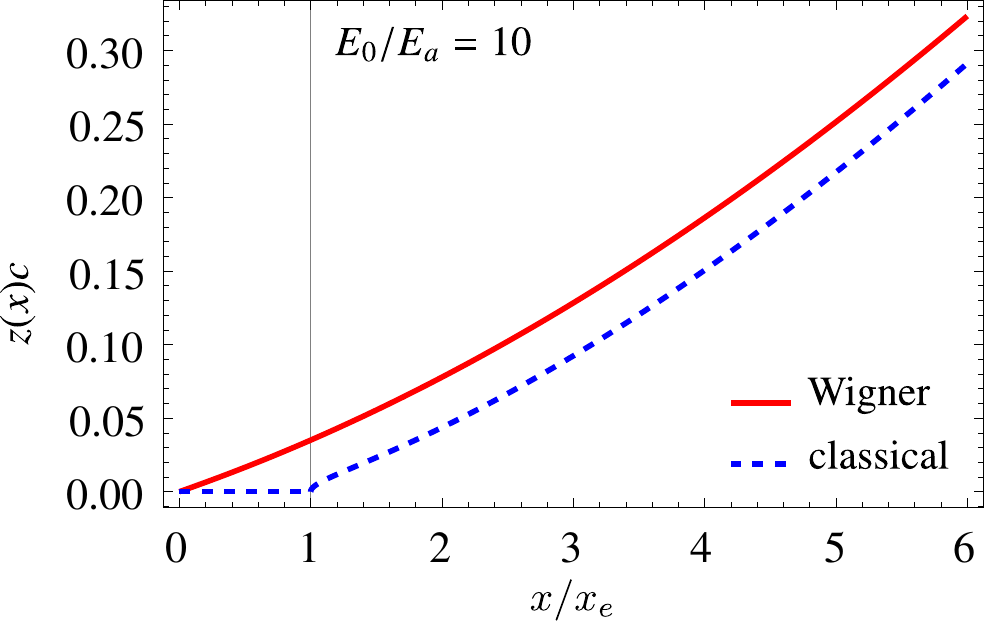}
  \caption{(Color online) Comparison of the Wigner and the classical trajectory in terms of the
  transversal coordinate $z$ for the parameters $E_0/E_a=10$ and $\kappa=90$.}
  \label{drift_cross_fields}
\end{figure}

For the comparison of the Wigner trajectory and the classical trajectory, we need to evaluate
the classical equations of motion. Let us calculate the classical trajectory for a relativistic
particle via the proper time parametrization $\tau$. The classical equations of motion are governed
by
\be
\label{Lorentz_force}
{\ddot{x}(\tau)}^\mu = - \frac{1}{c} {F^\mu}_\nu (x) {\dot{x}(\tau)}^\nu  \, 
\ee
with the field strength tensor $F_{\mu \nu}$.

For a constant and uniform crossed field~(\ref{crossed_fields}), the solutions are given by
\bal
\frac{1}{c} x^0 (\tau) & = \frac{\tau}{6 c^3}\sqrt{c^2+{v_z}_0^2} \left(6 c^2+E_0^2 \tau
^2\right) - \frac{E_0^2 \tau ^3 {v_z}_0}{6 c^3} \, ,\\
x^1 (\tau) & = \frac{E_0 \tau ^2}{2 c} \left({v_z}_0-\sqrt{c^2+{v_z}_0^2}\right) +x_e \, , \\
x^3 (\tau) & = \frac{E_0^2 \tau ^3}{6 c^2}\left(\sqrt{c^2+{v_z}_0^2}-{v_z}_0\right) + \tau
{v_z}_0
\eal
with the initial conditions $x^\mu (0) = (0,x_e,0)$ and $\dot{x}^\mu (0) = \left(\sqrt{c^2 +
{v_z}_0^2},0,{v_z}_0 \right)$, where we used the Lorentz invariant relation for the four-velocity
$\dot{x}^\mu \dot{x}_\mu = c^2$ \footnote{In principle, one can eliminate the proper time $\tau$ and
write down the equation of motion in terms of the parameter time $t$ as $x(t)$ and $z(t)$.}.
Moreover, the initial velocity along the propagation direction of the crossed field can be written
as ${v_z}_0 = c \,  q_z (x_e) /\sqrt{c^2 + q_z (x_e)^2}$ with $q_z (x_e) = p_z - x_e E_0/c $ and the
tunnel exit is given by 
\be
x_e = - \frac{I_p}{E_0} \left( \frac{18 c^2 - 5 I_p}{18 c^2 - 6 I_p} \right) \, ,
\ee
where we have used $p_z = - 2 I_p /(3c)$ in Eq.~(\ref{tunnel_exit_crossed_field}).

Similarly to the previous case, we can compare the Wigner trajectory with the classical trajectory
in two distinct regimes, see Fig.~\ref{time_delay_cross_fields_fixpro}. In the deep tunneling
regime, $E_0/E_a=1/7$, the Wigner time delay vanishes. For the near-threshold-tunneling regime,
$E_0/E_a=1$, the Wigner time delay is detectable.

Furthermore, one can compare the Wigner and the classical trajectories in terms of the
transversal coordinate $z$ (path). The stationary phase condition identifies the transversal
coordinate as
\be
\label{wigner_traject_via_fix_energy_prop_z}
z_\mathrm{TI} (x) \equiv - \left. \frac{\del \phi(x,\varepsilon_0, p_z)}{\del
p_z} \right|_{p_z = {p_z}_0} + \left. \frac{\del \phi(0,
\varepsilon_0,p_z)}{\del p_z} \right|_{p_z = {p_z}_0}\,
\ee
with the most probable momentum ${p_z}_0 = - 2 I_p / (3 c)$. Due to the existence of a non-zero
Wigner time delay, there exists also a spatial drift along the propagation direction at the tunnel
exit, which is also detectable at remote distance, see Fig.~\ref{drift_cross_fields} for an extreme
parameter $E_0/E_a = 10$.

\section{Conclusion} \label{conc}

In this brief report, we present an alternative method for calculation of the Wigner trajectory
which employs the fixed energy propagator. The developed method is applied to nonrelativistic as
well as relativistic tunnel-ionization from a zero-range potential, where the propagator is derived
from the proper-time method in the latter regime. We compare the quasiclassical trajectory with the
Wigner trajectory for the ionization processes in the deep-tunneling and the
near-threshold-tunneling regime. It is shown that the Wigner time delay is detectable in the latter
case and the results are in accordance with those of Ref.~\cite{Yakaboylu_2013_rt}.

\section*{Acknowledgments}

We are grateful to C. H. Keitel, S. Meuren, and A. W\"{o}llert for valuable discussions.

\bibliography{yakaboylu_bibliography}

\end{document}